\documentclass[9pt,twocolumn,twoside]{opticajnl}
\journal{opticajournal} 
\usepackage{hyperref}
\usepackage{amsmath}
\setboolean{shortarticle}{true}

\graphicspath{{Figures/}}
\usepackage{lineno}

\title{Three-Octave Supercontinuum Generation in Thick Crystalline Aluminum Nitride Waveguides}

\author[1,*]{Samantha Sbarra}
\author[2]{Samuele Brunetta}
\author[1]{Pierre Arnaud Demongodin}
\author[2]{Jean-François Carlin}
\author[2]{Nicolas Grandjean}
\author[2]{Raphaël Butté}
\author[1]{Camille-Sophie Brès}

\affil[1]{Ecole Polytechnique Fédérale de Lausanne, Photonic Systems Laboratory (PHOSL), CH-1015, Switzerland}
\affil[2]{Ecole Polytechnique Fédérale de Lausanne, Laboratory of Advanced Semiconductors for Photonics and Electronics (LASPE), CH-1015, Switzerland}

\affil[*]{samantha.sbarra@epfl.ch}

\begin{abstract}
We report an efficient extension of supercontinuum generation through dispersion engineering of crystalline aluminum nitride (AlN)-on-sapphire waveguides. Using a tailored epitaxial regrowth of AlN epilayers and an optimized fabrication protocol, the dispersion sensitivity to the waveguide cross-section was enhanced allowing for a significant reach extension of both short and long dispersive wave with optimized pumping conditions, reaching down to 550 nm in the visible and up to 4.5 µm in the mid-infrared. 
\end{abstract}

\setboolean{displaycopyright}{false} 
\begin{document}
\maketitle
\footnotetext{Copyright 2025 Optica Publishing Group.
This is the author's accepted manuscript of the article Opt. Lett. 50, 7147-7150 (2025).
One print or electronic copy may be made for personal use only. Systematic reproduction and distribution, duplication of any material in this paper for a fee or for commercial purposes, or modifications of the content of this paper are prohibited.}
Supercontinuum generation (SCG) is a powerful technique for producing new optical frequencies, including in unconventional spectral regions, by exploiting the nonlinear interaction between a single pulsed laser and a material, without requiring a resonant structure or temporal synchronization.
The increased capabilities of microtechnology have made it possible to achieve a high level of control and tunability of spectral broadening of on-chip integrated supercontinuum sources.
The strong mode confinement and the wide variety of high nonlinear refractive index materials available for integration are central ingredients for the future implementation of compact and scalable broadband sources, for applications in absorption spectroscopy \cite{Grassani2019, Tagkoudi2020}, frequency comb self-referencing \cite{Carlson2017}, microscopy \cite{Mansour2007}, imaging \cite{Poudel2019, Hartl2001}, and biological-chemical and environmental sensing \cite{Mizaikoff2013}. 
A plethora of examples of SCG on-chip can be found in literature, reporting hundreds of terahertz of spectral broadening on different platforms such as silicon, silica, silicon-germanium, silicon nitride, lithium niobate on insulator, silicon carbide, and III-V semiconductors such as aluminum nitride, gallium nitride, or aluminum gallium arsenide \cite{Bres2023}.
The key properties that influence a material's performance in SCG are its transparency window, the magnitude of its $\chi^{(2)}$ and $\chi^{(3)}$ nonlinearities, its linear refractive index dispersion, and its technological maturity, including flexibility in dimension design, availability of substrate and cladding materials, and ease/quality of fabrication.
Among the materials previously listed, aluminum nitride (AlN) exhibits highly attractive properties for SCG applications, in particular its exceptionally wide transparency window extending down to 200~nm, enabled by its 6.2~eV direct band gap \cite{Yim1973}. With its $\sim$ 2.1 refractive index at 1550~nm \cite{Bowman2018}, it provides substantial index contrast with standard silica cladding ($n\approx$ 1.5) and substrate materials, such as sapphire ($n \approx$ 1.75) or silica. Propagation losses can be as low as 0.1~dB/cm \cite{Liu2022} and 0.16~dB/cm \cite{Singh2024} depending whether crystalline or polycrystalline AlN is used, respectively. 
Along with a Kerr coefficient ($n_2$) of 2.3 $\times$ 10$^{-15}$~cm$^2$/W \cite{Jung2013}, AlN non-centrosymmetric nature provides an intrinsic second-order nonlinearity enabling three-wave mixing processes and f-to-2f interferometry.

Previous SCG realizations reported broadening from 0.5 to 4~µm in a SiO$_2$ clad 800~nm-thick polycrystalline AlN waveguide (WG) \cite{Hickstein2017} and from 0.7 to 3.5~µm \cite{Lu2020OL} on silica-clad crystalline AlN-on-sapphire 1~µm-thick WGs, both with a TE-polarized 1550~nm pump.
Although UV coverage has been achieved by chirping the WG and pumping at $\sim$800~nm \cite{Liu2019} or by exploiting the modal dispersion of high-order modes \cite{Chen2021a}, extending the reach of the SCG to the long-wavelength side of the AlN transparency window, while maintaining a good short wavelength coverage remains challenging.
Long wavelength reach could theoretically extend up to 13.6~µm \cite{Morkoc2009}, but in practice, it is limited by the absorption of the sapphire substrate and the silica cladding, when the optical mode is less confined in the WG core.

Previous works \cite{Grassani2019, Tagkoudi2020} achieved broadband mid-IR SCG in SiN by combining large cross-section WGs fabricated using the Damascene process, which yields thick, crack-free layers, and fiber laser pumping near 2~µm, taking advantage of recent developments in thulium-doped fiber laser technology. Here, we apply a similar approach by employing an AlN regrowth technique to extend the dispersion engineering space by means of thicker WGs. This not only enables the required dispersion control but also leverages the superior material transparency and intrinsic second-order nonlinearity of AlN, resulting in improved supercontinuum extension and efficiency, both towards the mid-IR and the visible range, with dispersive waves (DWs) near 4500 nm and 550 nm.
Furthermore, the design of WGs with large cross-sections helps to reduce losses induced by the cladding and substrate materials, ensuring modal confinement over a wider wavelength range.
\begin{figure}[ht]
  \centering
  \includegraphics[width=0.8\linewidth]{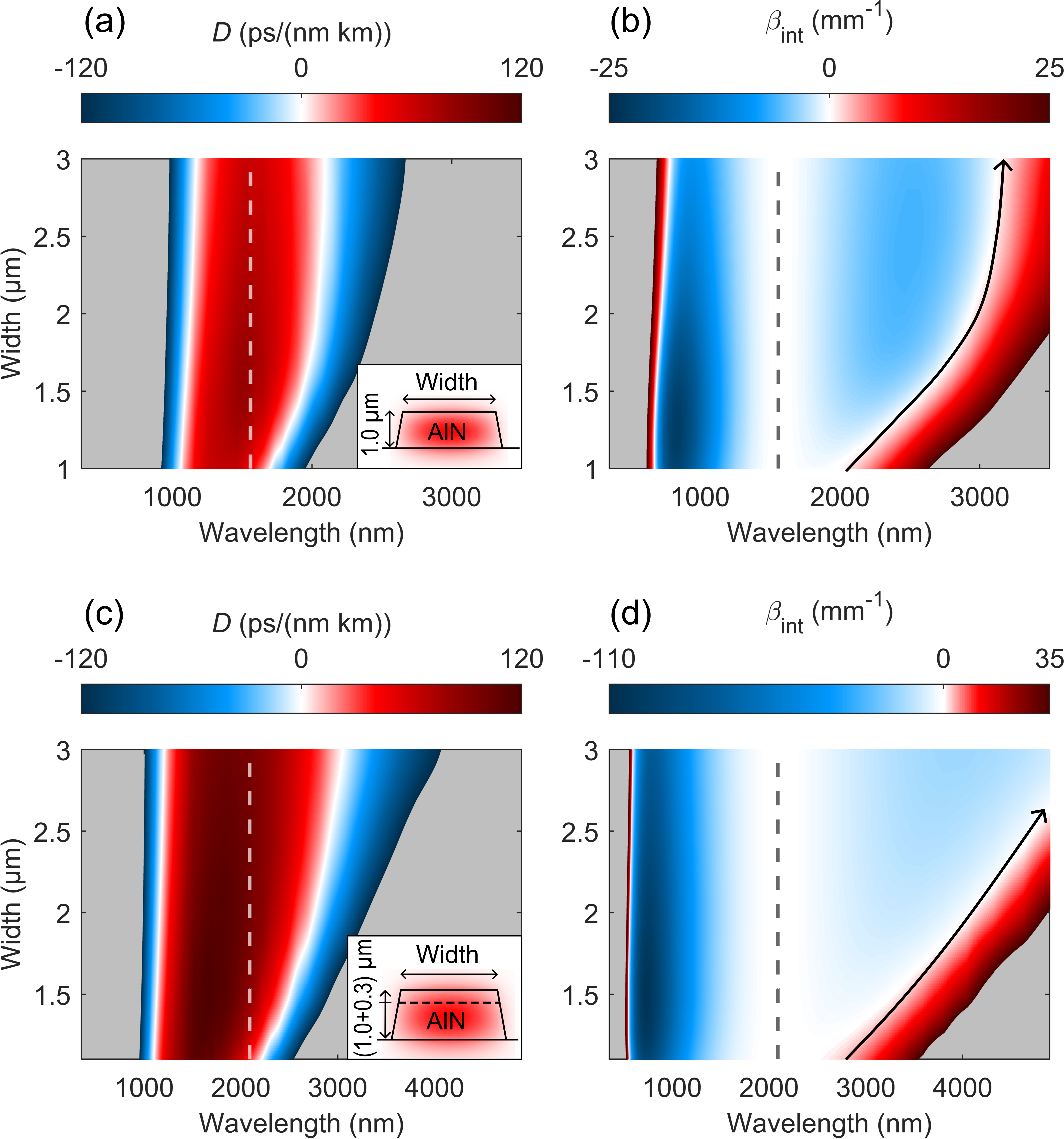}
  \caption{(a, b) Simulated group-velocity dispersion ($D$) and integrated dispersion ($\beta_\text{int}$) for a 1~µm-thick WG pumped at 1.56~µm (TM), and (c, d) for a 1.3~µm-thick WG pumped at 2.1~µm (TM). Insets in (a, c) show normalized mode profiles at the pump wavelength for 2.5~µm width; the 0.3 µm regrown layer is indicated by a dashed line in (c) inset. Normal (anomalous) dispersion is shown in blue (red); out-of-range values appear in gray. Dashed lines: pump position used for $\beta_\text{int}$; black arrows: LWDW shift with WG width.}
  \label{fig:Figure1}
\end{figure}
While the width of a WG is determined by the lithographic process used to transfer the chip design onto the material substrate and can easily accommodate micrometer scale dimensions, the height is defined by the growth process: mostly epitaxial techniques for crystalline materials and deposition techniques, e.g., sputtering for amorphous materials. For both types of materials care has to be taken to remain below the plastic relaxation limit and hence avoid the appearance of crippling extended defects such as cracks and/or structural defects.
In addition to its superior optical performance compared to its polycrystalline counterpart obtained by sputtering, crystalline AlN can be grown to larger thicknesses without incurring a surface roughness that could otherwise compromise its overall optical qualities and was shown to be less prone to photon absorption in the UV and mid-IR ranges \cite{Stegmaier2014, Liu2018a, Liu2018}.
We started our analysis on WGs of 1~µm thickness, corresponding to the typical value of commercially-available crystalline AlN epilayers grown on \textit{c}-plane sapphire substrate.
First, the effective refractive index ($n{_\text{eff}}$) and modal dispersion for different cross-sections were calculated using finite-element modal analysis.
For this numerical study, the extraordinary and ordinary refractive indices of AlN were computed from the Sellmeier equation proposed in \cite{Soltani2014}.
Figure \ref{fig:Figure1}~(a) shows the resulting group velocity dispersion, $D$ = -(2$\pi c_0$)/$\lambda^2 \times \beta_2$, for different WG widths in TM polarization. Here, $c_0$ stands for the velocity of light in vacuum, $\beta_2$ for the second derivative of the wavenumber $\bar{k}$ with respect to the angular frequency $\omega$.
In this work, we focused on TM optical modes, anticipating future interest in the self-referencing of SC spectral lines (results in the TE polarization can be found in the Supplement 1).
This requires efficient pump second-harmonic generation (SHG) through excitation of a TM mode aligned with the extraordinary axis, along which the largest nonlinear tensor component ($\chi^{(2)}_\text{33}$) of AlN is oriented \cite{Chen1998,Majkic2017}.
The blue and red regions in Figure \ref{fig:Figure1}~(a) indicate the normal and anomalous dispersion regions, respectively. SCG based on soliton dynamics requires excitation of the WGs in the anomalous region, and the optimal excitation wavelength was found around 1.56~µm (dashed line), corresponding to the largest value of $|D|$. This condition ensures the largest broadening of the SC spectrum with well separated DWs, and a short fission length, without compromising the number of generated solitons \cite{Bres2023}. 
Therefore, the integrated dispersion with respect to a 1.56~µm pump is calculated as $\beta_\text{int}(\omega)=\beta(\omega)-\beta(\omega_0)-(\omega - \omega_0)/v_\mathrm{g}$, where ${\beta(\omega)}$ is the frequency-dependent propagation constant of the optical mode, and $v_\mathrm{g}$ is the group velocity at the pump angular frequency $\omega_0$. The estimated spectral position of the DWs is found in correspondence of the phase-matching (PM) conditions $\beta_\text{int}=0$.
The blue- and red-detuned DWs are referred to short-wavelength dispersive wave (SWDW) and long-wavelength dispersive-wave (LWDW), respectively.  
From the plot of $\beta_\text{int}$ in Figure \ref{fig:Figure1}~(b), we can establish that for a 1~µm-thick WG the position of the SWDW lies between 680 and 780~nm, and is only slightly influenced by the WG width owing to the strong dependence of the chromatic dispersion.
The LWDW, on the other hand, red-shifts from 2 to 3~µm as the WG width increases from 1 to 2~µm. Beyond this value, its position remains nearly unchanged, as the dispersion becomes insensitive to further increases in width (black arrow).
Pushing the LWDW further into the infrared requires the use of thicker WG layers. 
Figure \ref{fig:Figure1}~(c) shows that by increasing the WG thickness to 1.3~µm the anomalous dispersion region extends toward longer wavelengths, accompanied by a favorable increase in $|D|$. The optimal pumping condition is thereby shifted towards 2~µm (dashed line), is compatible with commercial fiber lasers and integrated thulium-based sources \cite{Su2016, Li2017b}.
The $\beta_{\text{int}}$ shown in Figure \ref{fig:Figure1}~(d) reveals a blue shift of the SWDW down to approximately 550~nm. At the same time, the position of the LWDW is strongly influenced by the WG width over a broader range, ultimately extending beyond 4~µm.

\begin{figure}[ht]
  \centering
  \includegraphics[width=0.9\linewidth]{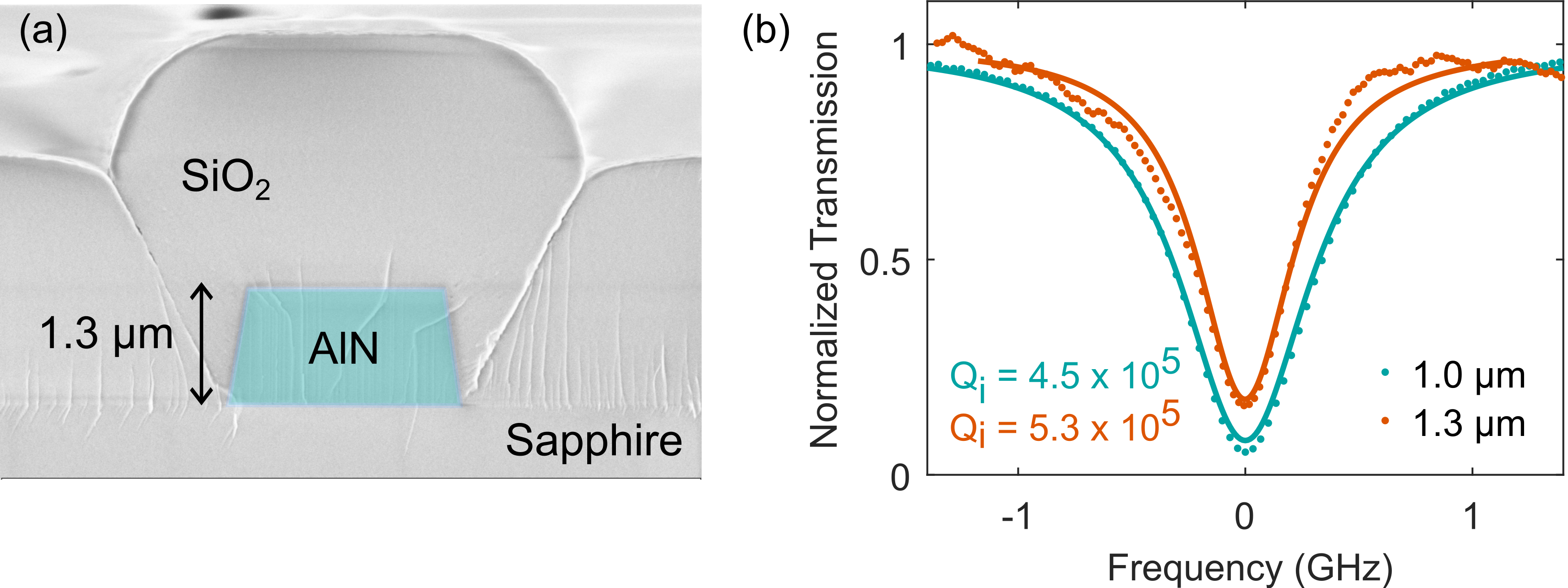}
  \caption{(a) False-color SEM image of an AlN WG facet from a regrown epilayer with silica cladding. (b) TM resonances at 1550~nm from microrings fabricated on 1.0~µm- and 1.3~µm-thick epilayers. The corresponding $Q_\text{i}$ are indicated.}
  \label{fig:Figure2}
\end{figure}

Following our numerical results, we implemented an in-house fabrication process to produce crystalline AlN WGs on sapphire substrates with thicknesses not commercially available.
In this article, we investigate WGs fabricated from a commercially available 1~µm-thick AlN epitaxial layer and compare them with those issued from epilayers involving an MOVPE regrowth step to increase the initial thickness of the commercial epitaxial layer (1~µm) up to a total thickness of 1.3~µm, while preserving the initial surface quality (see the corresponding atomic force microscopy image in Supplement 1).
For both types of epilayers, the WG geometry was defined by an e-beam lithography step using a thick negative-tone FOx-16 hydrogen silsesquioxane resist deposited on a 30~nm titanium anti-charging layer. 
To achieve full-depth WG etching, enhancing mode confinement and preventing slab leakage, the lithographic mask must be precisely defined, as the AlN-to-mask etch selectivity is slightly above 2.
The inductively-coupled plasma etching step was based on a Cl\textsubscript{2}-BCl\textsubscript{3}-Ar gas mixture and produced $\sim$80$^\circ$ slanted sidewalls.
The sample was clad with silica using plasma-enhanced chemical vapor deposition. Finally, the sample facets were dice-and-cleaved \cite{Chen2014}. 
A scanning electron microscopy (SEM) image of a 2.3~µm $\times$ 1.3~µm WG facet is shown in Figure \ref{fig:Figure2}~(a).

Propagation losses ($\alpha$) were derived from the intrinsic quality factors ($Q_i$) of 60~µm-radius, 1.2~µm-wide ring resonators from both epilayers and measured with a 1550~nm TM pump. 
Low-power transmission spectra are shown in Figure \ref{fig:Figure2}~(b). For the 1.0~µm and the 1.3~µm-thick WGs, $Q_\text{i}$ values of 4.5 $\times 10^5$ ($\alpha\approx$0.8~dB/cm) and 5.3 $\times 10^5$ ($\alpha\approx$0.7~dB/cm) were obtained, respectively, with sidewall roughness likely being the limiting factor. These results indicate that the regrowth process does not degrade propagation losses.

\begin{figure}[ht]
  \centering
  \includegraphics[width=0.9\linewidth]{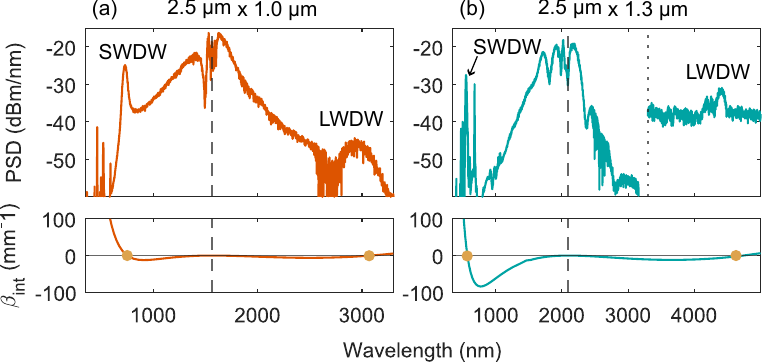}
  \caption{OSA SC spectra from a 2.5~µm $\times$ 1~µm WG with a TM 1.56~µm pump (a) and from a 2.5~µm $\times$ 1.3~µm WG with a TM 2.1~µm pump (b). Dashed line: pump position. Simulated $\beta_\text{int}$ shown below, with SWDW and LWDW positions indicated by yellow dots. In (b), an FT-OSA with a -40~dBm/nm noise floor is used beyond 3.3~µm (dotted line).}
  \label{fig:Figure3}
\end{figure}

Our chips were characterized using two femtosecond lasers, one centered at 1.56~µm (TOPTICA FemtoFErb) with a repetition rate of 100~MHz and a pulse width of 46~fs, and one at 2.1~µm (NOVAE $\lambda+$) operating at a repetition rate of 20 MHz, with a pulse width of 79 fs.
A polarization stage allowed selection between TE and TM polarizations, and an aspheric lens was used to couple light into our 4.5~mm long WGs.
We measure an average power of 16.7~dBm (pulse energy of $\sim$ 460~pJ) for the 1.56~µm laser, and 17.8~dBm ($\sim$ 3~nJ) for the 2.1~µm laser at the input of the coupling optic.
Depending on the quality of the facet cleavage and the WG width, coupling losses can vary from 2 to 5~dB/facet. As such, the actual coupled power is not trivial to determine, and for the sake of consistency, we state the measured power value before the coupling objective rather than that estimated on-chip.
Collection of the outcoupled light is done with an achromatic objective, followed by a collimator and a multimode fiber.
Two optical spectrum analyzers (OSAs) were employed to cover the visible and IR range, while a Fourier-transform optical spectrum analyzer (FT-OSA) was used to detect the signal from 3.3 to 5.5~µm, with a $\sim$-40~dBm/nm noise floor. The setup schematic is shown in Supplement 1.

The OSA power spectral density (PSD) of the SC spectra obtained from two different experiments are presented in Figure \ref{fig:Figure3}. 
The spectrum reported in Figure \ref{fig:Figure3}~(a) was collected from a 2.5~µm-wide WG fabricated on the 1~µm-thick commercial epilayer pumped at 1.56~µm. The spectrum extends from around 700~nm to 3~µm, in agreement with the theoretical positions of the DWs shown in the bottom panel.  
However, the signal at the SWDW is approximately 20~dB stronger than at the LWDW, because the broadened soliton (located near the pump wavelength) has greater spectral overlap with the former, facilitating a more efficient energy transfer to shorter wavelengths. 
These results are consistent with those reported in \cite{Lu2020OL} on WGs of similar cross-section but 8~mm long and an estimated on-chip pulse energy of 700~pJ. 
Owing to the expected normal dispersion at 2~µm in this WG (see Figure \ref{fig:Figure1}~(a)), we do not anticipate any significant broadening beyond self-phase modulation.
We then tested a WG of the same width fabricated from the "regrown" epilayer of $\sim$1.3~µm thickness. 
Pumping this WG at 1.56~µm led to an inefficient energy transfer towards long wavelengths due to the large integrated dispersion and the small spectral overlap with the soliton (see Supplement 1 for the corresponding SC spectra).
The long wavelength reach and amplitude of the SC signal increase significantly when the 2.1~µm pulsed laser is used, as visible in Figure \ref{fig:Figure3}~(b). 
In agreement with the numerical calculations anticipating a smaller value for $\beta_\text{int}$ in the long wavelength range (bottom panel), the separation between the DWs reaches more than three octaves, spanning from 550~nm to 4.5~µm. This exceptional broadening of the SC is accompanied by high power levels for both DWs, which are less than 10 dB below the pump, well above the noise floor of the FT-OSA.
Additional features can be identified in both spectra, which further extend the SC towards the short-wavelength range. These are related to four-wave mixing processes such as third-harmonic generation (THG), and $\chi^{(2)}$ processes such as SHG, cascaded SHG to the 4$^{\text{th}}$ harmonic, and sum-frequency generation (SFG), involving the PM higher-order modes supported by the multimode WG. Furthermore, absorption lines from CO$_2$ and water-vapor molecules from the environment become visible around 2.7~µm.

\begin{figure*}[ht]
  \centering
  \includegraphics[width=\linewidth]{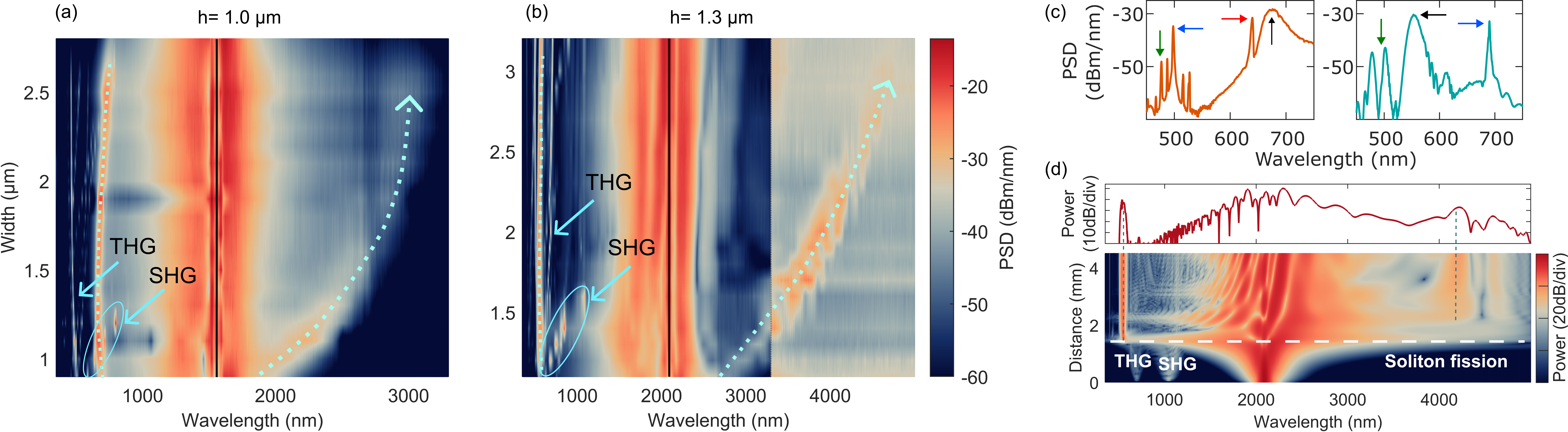}
  \caption{Surface plots of measured SC spectra from WGs of different widths and 1~µm (a) and 1.3~µm (b) thickness, pumped with TM polarization at 1.56~µm and 2.1~µm, respectively. DWs are indicated by the dotted lines; SHG and THG peaks by arrows. (c) PSD at short wavelengths for the 1~µm$\times$1~µm (left) and 2.5~µm$\times$1.3~µm (right) WGs, with arrows indicating the SWDW (black), SHG (red), THG (blue) and 4$^\mathrm{th}$ harmonic (green). (d) Simulated SC spectrum (top) and pulse spectral broadening along the propagation direction (bottom) for the 2.5~µm $\times$ 1.3~µm WG pumped at 2.1~µm.}
  \label{fig:Figure4}
\end{figure*}
We characterized the SCG from WGs with different widths to test the sensitivity of the dispersion to the WG cross-section and report the results in Figure \ref{fig:Figure4}~(a, b). 
In agreement with the numerical results (Figure \ref{fig:Figure1}), the SC spectra show an efficient generation of the SWDW in the 1~µm-thick WGs and a weaker LWDW, whose extension is limited to 3~µm.
Instead, in 1.3~µm-thick WGs, both DWs are efficiently generated, with the LWDW continuously red-shifting with increasing widths. 
The calculation of the efficiency of DWs generation for both WG thicknesses is presented in Supplement 1.
The evolution of the PM condition for the signal generated through SHG and THG processes in both plots is also indicated. The intensity of the signal stemming from these processes is shown in more detail in Figure \ref{fig:Figure4}~(c) for the 1~µm-thick  (left panel) and the 1.3~µm-thick (right panel) epilayers, highlighting the enhanced emission in the visible.
Higher-order modes, although supported for large cross sections, are not excited due to their spatial symmetry and/or exhibit high losses as they are close to cutoff.

Simulations of the pulse broadening in the 2.5~µm x 1.3~µm WG pumped at 2.1~µm were carried out employing the open-source \textit{Python} package provided in \cite{Voumard2023}. These numerical simulations are suited for materials possessing both quadratic and cubic nonlinearities and were modified to account for the wavelength-dependent modal effective area. 
The bottom panel of Figure~\ref{fig:Figure4}~(d) shows the spectral evolution of the pulse along the propagation direction. While phase-unmatched SHG and THG appear at the beginning of the propagation, the SC broadens gradually and soliton fission occurs at around 1.5~mm distance from the WG input.
The resulting output spectrum shown in the top panel is in good agreement with the one obtained experimentally and shown in Figure \ref{fig:Figure3}~(b) and the position of the DWs is consistent with the one observed on the (FT-)OSA. The amplitude of the LWDW appears slightly attenuated in the experiments compared to that expected from the simulations. We attribute this discrepancy to an increase in absorption and scattering in the cladding that plays a non-negligible role when the confinement decreases at longer wavelengths.
At short wavelengths, the amplitude of the SWDW is consistent with the one measured but no additional peaks from PM non-linear processes involving higher-order modes are retrieved because only the fundamental mode is accounted for in our simulations. Simulated and experimental SC spectra for both WGs shown in Figure \ref{fig:Figure3} are compared in Supplement 1.

In conclusion, we have demonstrated three-octave bandwidth supercontinuum generation from crystalline AlN WGs on sapphire, with the specific goal of efficiently transferring the pulse energy towards the mid-IR region up to 4.5~µm, foreseeing future applications in the field of spectroscopy. This was achieved on the basis of flexible selection of the WG dimensions provided by in-house growth and fabrication facilities. To our knowledge, this three-octave bandwidth spectrum is the widest achieved with an IR pump on materials possessing both quadratic and cubic nonlinear responses such as lithium niobate \cite{Yu2019, Lu2019OL, Fiaboe2024}, aluminum gallium arsenide \cite{Kuyken2020} or indium gallium phosphide \cite{Dave2015}, that might offer the possibility of providing a comb spectrum and its frequency reference on the same chip simultaneously.  

\vspace{0.2cm}
\noindent{\textbf{Funding.}}
École Polytechnique Fédérale de Lausanne; Schweizerischer Nationalfonds zur Förderung der Wissenschaftlichen Forschung (200020\_215633).

\vspace{0.2cm}
\noindent{\textbf{Disclosures.}}
The authors declare no conflict of interest.

\vspace{0.2cm}
\noindent{\textbf{Data availability Statement}}
Data underlying the results presented in this paper are not publicly available at this time but may be obtained from the authors upon reasonable request.

\vspace{0.2cm}
\noindent{\textbf{Supplemental document}}
See Supplement 1.

\bibliography{NonLinearOptics}
\bibliographyfullrefs{NonLinearOptics}
\end{document}